\documentclass[aip,preprint,showpacs,amsmath,amssymb]{revtex4}

\usepackage{graphicx}
\usepackage{dcolumn}
\usepackage{bm}

\begin{document}
\title{Optimal electro-mechanical control of the excitonic fine structures of droplet epitaxial quantum dots}
\author{Shun-Jen Cheng}
\affiliation{Department of Electrophysics, National Chiao Tung University, 
Hsinchu 300, Taiwan, Republic of China}

\author{Yi Yang}
\affiliation{Department of Electrophysics, National Chiao Tung University, 
Hsinchu 300, Taiwan, Republic of China}

\author{Yu-Nien Wu}
\affiliation{Department of Electrophysics, National Chiao Tung University, 
Hsinchu 300, Taiwan, Republic of China}

\author{Yu-Huai Liao}
\affiliation{Department of Electrophysics, National Chiao Tung University, 
Hsinchu 300, Taiwan, Republic of China}

\author{Guan-Hao Peng}
\affiliation{Department of Electrophysics, National Chiao Tung University, 
Hsinchu 300, Taiwan, Republic of China}

\date{\today}

\begin{abstract}
The intrinsic fine structure splittings (FSSs) of the exciton states of semiconductor quantum dots (QDs) are known to be the major obstacle for realizing the QD-based entangled photon pair emitters.    
In this study, we present a theoretical and computational investigation of the excitonic fine structures of droplet-epitaxial (DE) GaAs/AlGaAs QDs under the electro-mechanical control of micro-machined piezoelectricity actuators. 
From the group theory analysis with numerical confirmation based on the developed exciton theory, we reveal the general principle for the optimal design of micro-machined actuators whose application on to an elongated QD can certainly suppress its FSS. 
We show that the use of two independently tuning stresses is sufficient to achieve the FSS-elimination but is not always necessary as widely deemed. The use of a single tuning stress to eliminate the FSS of an elongated QD is possible as long as the crystal structure of the actuator material is in coincidence with that of the QD.
As a feasible example, we show that a {\it single} symmetric bi-axial stress naturally generated from the $(001)$ PMN-PT actuator can be used as a single tuning knob to make the full FSS-elimination for elongated DE GaAs QDs.
 \end{abstract}


\maketitle

\section{Introduction}
Generation of polarization-entangled photon pairs is a vital element in the advanced quantum photonic applications, such as quantum cryptography and quantum teleportation. \cite{Entangle}
Semiconductor quantum dots was predicted to be a promising nano-material for being "on-demand" entangled photon pairs emitters (EPPEs), which are key devices necessary in quantum cryptography and quantum teleportation.\cite{Entangle1, Mueller, Young2006, Akopian}  
However, in reality photo-excited QDs usually fail to generate such polarization entangled photon pairs because of the intrinsic FSSs between the single bright-exciton doublet as the intermediate states in the process of spontaneous biexciton-exciton-vacuum cascade decay. The FSS of an exciton in an self-assembled QD is caused by the electron-hole ({\it e-h}) exchange interactions which are likely induced by any slight symmetry breakings of QD structures, such as shape elongation, strain, or composition randomness, and leads to the destruction of entanglement with the reveal of the which-path-information in the processes of spontaneous exciton decay.\cite{Zunger} 
Thus, technologies for fully eliminating the exciton FSSs, against the inherent or extrinsic symmetry breakings, of QDs have been desired for a long time and are still being under the active development for the QD-based photonic applications.\cite{Bennett2010,Mohan, Trotta, Kuroda}

In the earlier time, most experiments attempted to use single generic fields, e.g. electrical,\cite{Kowalik2005,Kowalik2007,Bennett2010} magnetic,\cite{Stevenson, Lin} optical,\cite{Muller} or stress fields \cite{Seidl}, as {\it single} tuning knobs to suppress the FSSs of QDs, but the yield of successful devices was extremely low.  As a known example, the FSSs of QDs can be well tunable by a single uniaxial stress but hardly really tuned to be zero.\cite{Seidl,Singh}

Till some years ago, a conceptual and technological breakthrough was first proposed and experimentally confirmed by Trotta {\it et al.} to solve the problem.\cite{Trotta} They show that at least {\it two} tuning knobs are needed for a thorough elimination of the FSSs of QDs.\cite{Trotta, Wang, Pooley}  In the experiment, with the simultaneous application of an uniaxial stress and vertical electrical bias, the FSSs of the inherently strained InAs/AlGaAs QDs were fully eliminated in a universal and deterministic manner. Years later, with the advances in the fabrication of micro-machined actuators, the deterministic generation of entangled photon pairs from InAs/AlGaAs QDs was also realized by means of simultaneously applying two independently tuning uni-axial stresses onto the QDs.\cite{Trotta2016}  The realization of the QD-based EPPEs by means of the electro-mechanical control opens up a prospect of the integration of QD-based photonics with micro-electro-mechanical systems (MEMSs). Inspired by the progress, currently more attempt is devoted to developing the versatile QD-based EPPEs with the functionalities useful in scalable integrated photonic systems.\cite{Trotta2016, Trotta2015, Chen2016} In the sense, the need and use of two tuning knobs for the FSS-tuning yet hinder the versatility of devices that also require additional tuning knobs for the functional operations.\cite{Zhang2015} 
Thus, for practical applications and also fundamental curiosity, the following questions arise: "Why two tuning knobs are necessary?" and "Can the number of required tuning knobs be reduced?"

In this work, we present a theoretical and computational investigation of the excitonic fine structures of inherently un-strained GaAs/AlGaAs DE-QDs under the electro-mechanical control, implemented by micro-machined  $[\rm{Pb(Mg}_{1/3}\rm{Nb}_{2/3})\rm{O}_3]_{0.72}\rm{-[PbTiO_3]}_{0.28}$ (PMN-PT) piezoelectricity actuators in the multi-legged structures.\cite{Trotta2015, Koguchi, Watanabe, Kumar2014} As compared with more extensively studied InAs/AlGaAs QDs grown in the Stranski-Krastanov (SK) mode,\cite{Bimberg, Michler} GaAs QDs grown by droplet epitaxial technique are advantageous in the well-controlled shape geometry \cite{Liao2012, Huber}, negligible inter-diffusion at interfaces \cite{Schlesinger}, and the absence of internal strain. \cite{Keizer} Notably, the absence of inherent strain makes the electronic and excitonic structures of GaAs DE-QDs sensitive to and highly tunable by external stresses. \cite{Kumar2014,Plumhof, Kumar2011, Sanchez}

From the group theory analysis with numerical confirmation based on the multi-band exciton theory, we derive explicitly the general principle for the optimal arrangement of uni-axial stresses from micro-machined PMN-PT actuators that can certainly suppress the FSSs of elongated GaAs DE-QDs fully. The principle to follow is that a full elimination of the FSS, tuned by external knobs, of an elongated QDs can be always possible as long as the symmetry of the QD can be kept invariant during the tuning process. Surprisingly, we find that that, beyond common intuitive understanding, the use of two tuning knobs is actually a sufficient but not a necessary condition for a deterministic elimination of the FSS of an elongated QD. 
As a feasible example, it is shown that a {\it single} symmetric bi-axial stress naturally generated from the $(001)$ PMN-PT actuator can be used as a single tuning knob to make the full FSS-elimination, and advantageous in the robustness of the FSS-tuning against the poorly controlled orientation variations of actuators. The deterministic elimination of the FSSs of zinc-blende GaAs QDs by using a single symmetric bi-axial stress from $(001)$ PMN-PT actuator is achievable by taking the advantage of the compatibility of crystal symmetry between the QD- and piezoelectricity actuator materials and essentially related to the stress-enhanced valence band mixing (VBM) in the QD-confined exciton.\cite{Luo}.  

This article is organized as follows. The next section presents the theoretical and computation methodology used throughout this work. In Sec.II-A, we present the group theory analysis for the excitonic fine structures of elongated semiconductor quantum dots under the action of generic tuning stresses. Sec. II-B is devoted to the theory of electron-hole exchange interaction in a QD-confined exciton and the numerical implementation for the simulation of the fine structures of stressed GaAs/AlGaAs QDs. In Sec. III, we discuss the predicted excitonic fine structures of stress-controlled QDs by the group theory. Then, we present a general principle for the optimal design of the stress actuators predicted from the analysis, with the confirmation by the numerical computation. In Sec. IV, we establish a valid simplified exciton model that incorporates the non-linear effect of the bi-axial stress and valence-band-mixed nature of exciton. Finally, we conduct the model analysis to discuss several advantageous features of the zinc-blende QDs controlled by a single symmetric bi-axial stress. Sec.V concludes this work.

\begin{figure}[t]
\includegraphics[width=15cm]{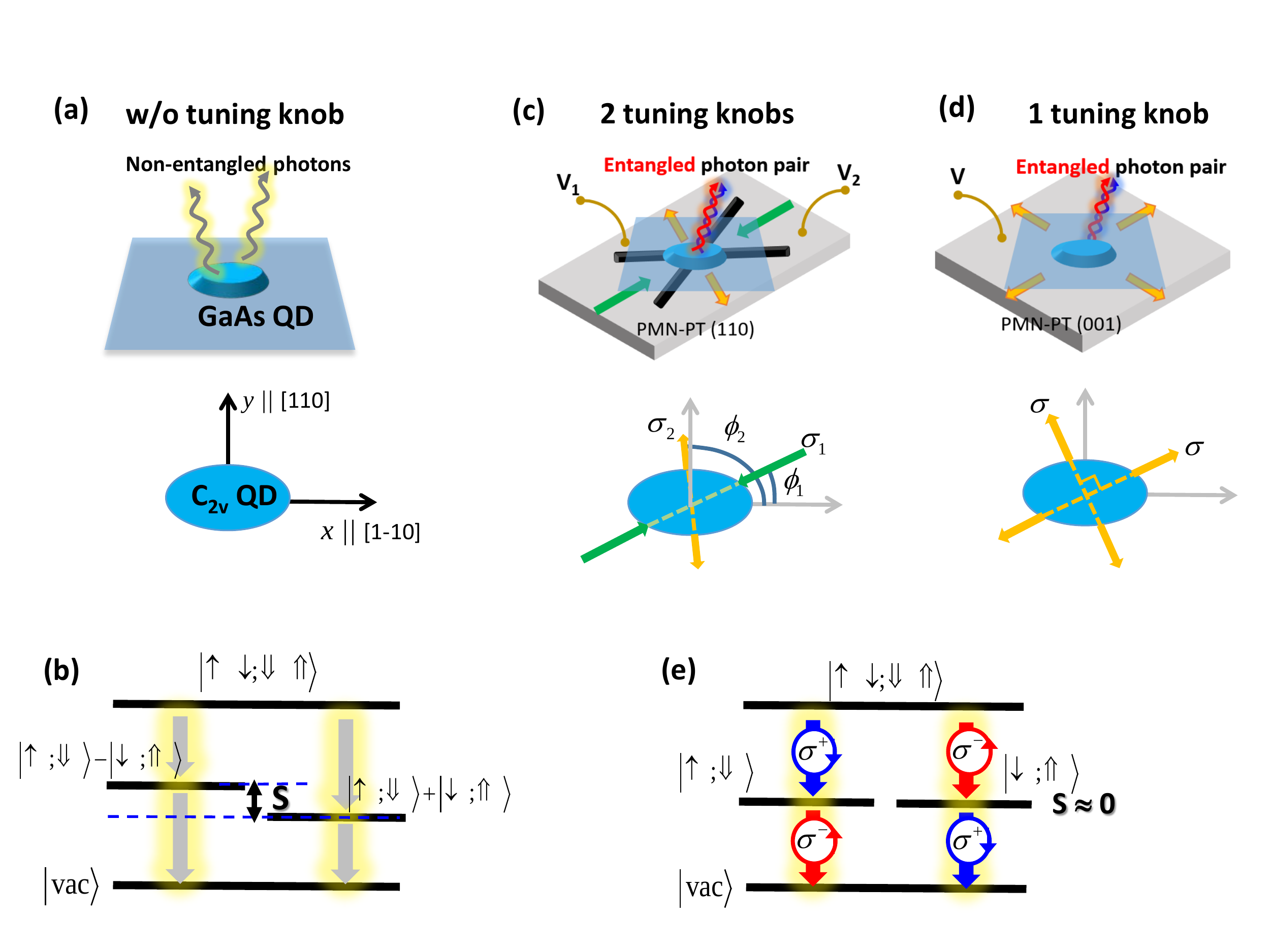}
\caption{
Schematics of (a) a photo-excited QD in the elongated shape of $C_{2v}$ symmetry that successively emits a pair of photons not in entanglement, and, correspondingly, (b) the bi-exciton and single exciton levels of the elongated QD, with a non-zero fine structure splitting ($S\neq 0$) between the single-exciton doublet. (c)  A $C_{2v}$ QD with two mechanical tuning knobs, a set of two uni-axial stresses generated and controlled by a $(110)$ micro-machined PMN-PT actuator.  Generation of polarization-entangled photon pairs from the stressed QD is possible if the strengths, $(\sigma_1,\sigma_2)$, and the orientations, $(\phi_1,\phi_2)$, of the two tuning uni-axial stresses are chosen appropriately. (d)  The $C_{2v}$ QD with a single tuning knob of symmetric bi-axial stress from $(001)$ PMN-PT crystal that can emit a pair of photons in entanglement with only appropriate adjustment of the strength of the single stress, irrelevant to the orientation of the stress-axes. (e) The exciton-level schematics of the QD that can emit entangled photon-pairs, where the fine structure splitting $(S)$ is intrinsically zero or eliminated by external tuning knobs.}
\label{Fig1}
\end{figure} 

\begin{figure}[t]
\includegraphics[width=15cm]{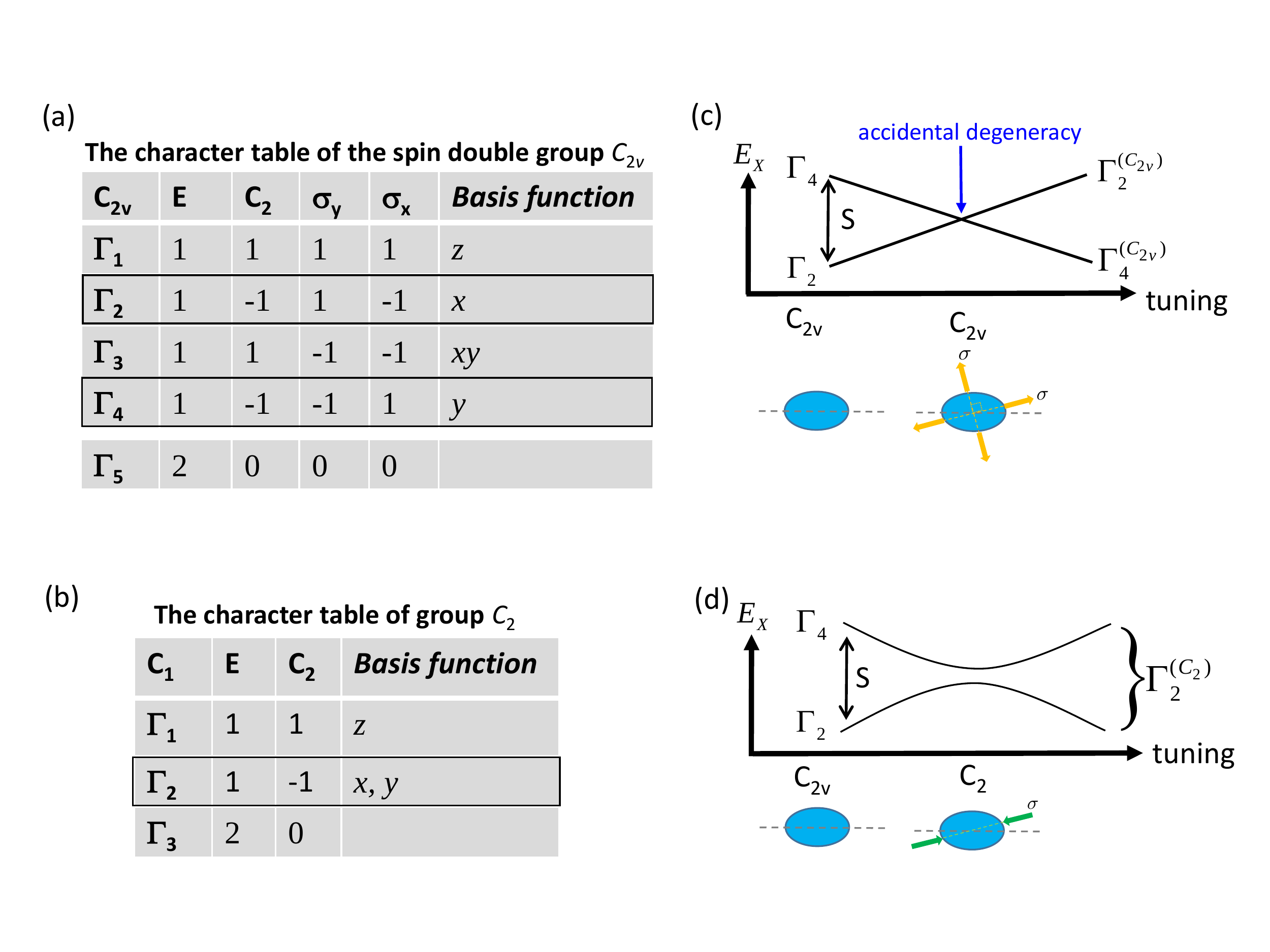}
\caption{
The character tables of (a) the spin double group $C_{2v}$ and (b) group $C_{2}$. (c) Schematics of the stress-tuned exciton levels of a $C_{2v}$ QD with the preservation of the $C_{2v}$ symmetry, e.g. by means of symmetric bi-axial stress. In the case, a direct crossing (leading to $S=0$) of the distinct exciton levels belonging to different irreducible representations, $\Gamma_2$ and $\Gamma_4$, can be made by an accidental degeneracy. (d) Schematics of the exciton levels of an elongated $C_{2v}$ QD whose symmetry is reduced to $C_{2}$ by a mis-aligned uni-axial stress from the elongation axis.  Without the preservation of the $C_{2v}$ symmetry, the two exciton states belong to the same irreducible representations, $\Gamma_2$, and no level crossing can happen. }
\label{Fig2}
\end{figure}

\begin{figure}[b]
\includegraphics[width=15cm]{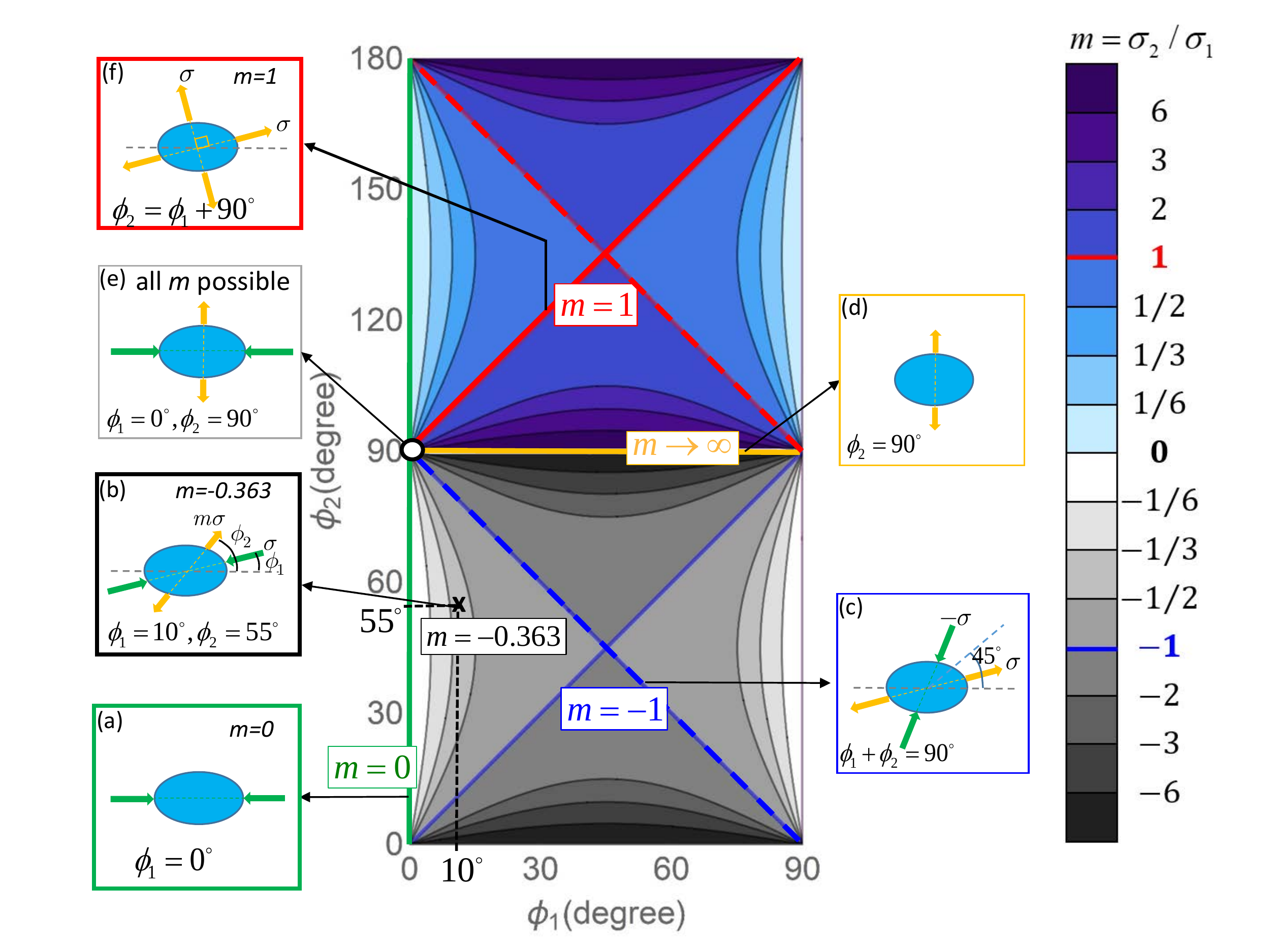}
\caption{
The contour plot of the stress ratio $m\equiv \sigma_2/ \sigma_1 $ for a set of two uni-axial stresses, as a function of the angles of the stress axes, $\phi_1$ and $\phi_2$, that fulfills Eq.(\ref{S0condit}) and allows for the possibility of mechanically tuning and making $S=0$. The insets depict various predicted useful dual-stress actuators that can deterministically make the FSS-elimination for $C_{2v}$ QDs.}
\label{Fig3}
\end{figure}

\begin{figure}[t]
\includegraphics[width=18cm]{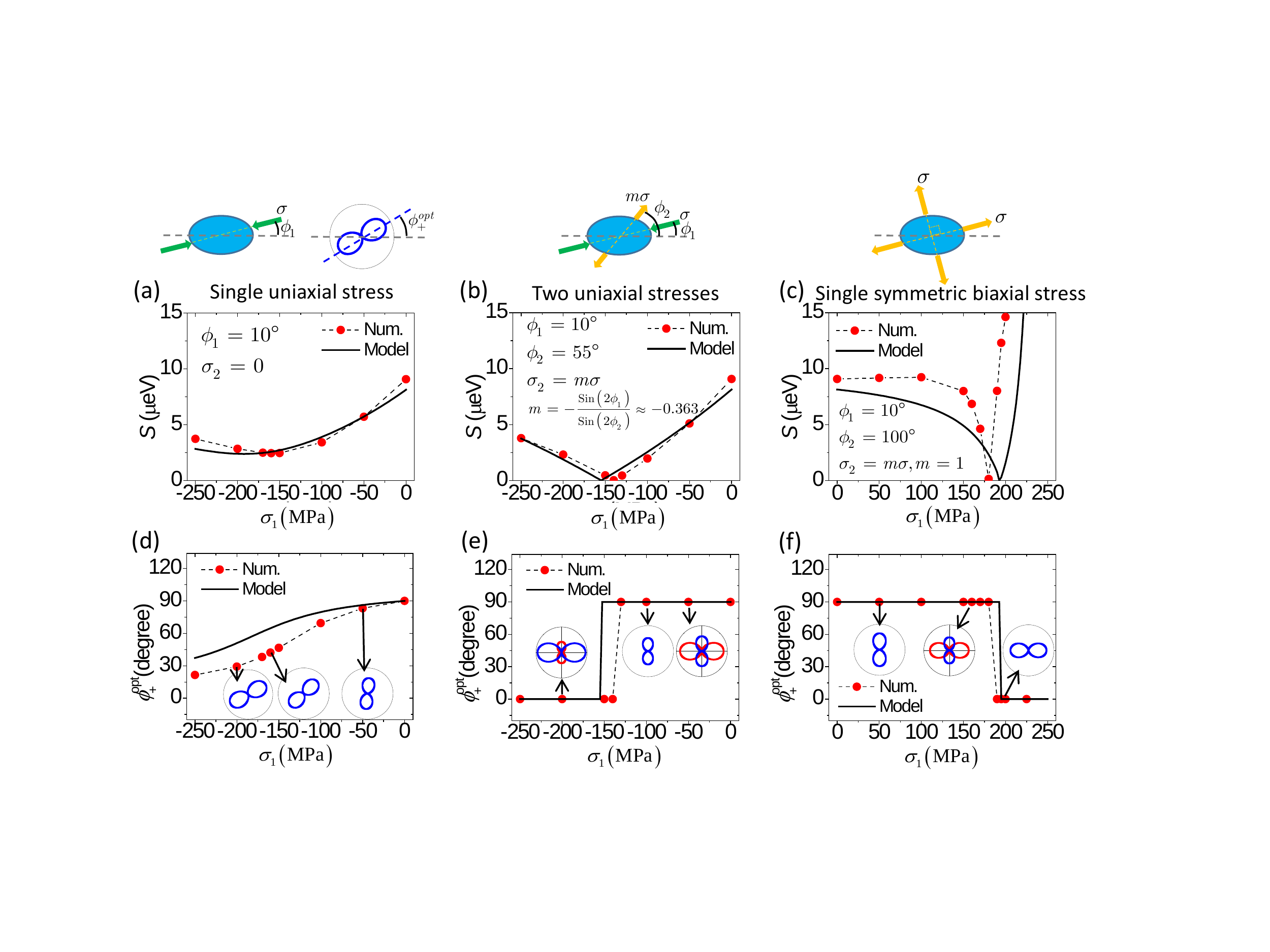} 
\caption{
The calculated fine structure splitting of the exciton doublet of a GaAs DE-QD under (a) an uni-axial stress misaligned to the elongation axis by the angle $\phi_1=10^\circ$, (b) a set of two uni-axial stresses with $\phi_1=10^\circ$, $\phi_2=55^\circ$ and the fixed stress ratio $m=\sigma_2/\sigma_1= \sin 2\phi_2/\sin 2 \phi_1= -0.363$, and (c) a single symmetric bi-axial stress. The numerical (model calculated) results are indicated by filled red circles (solid lines). (d)-(f): The angles, $\phi_{+}^{\rm{opt}}$, with respect to the QD-elongation axis, of the optical polarization axes of the upper level state of the exciton doublet of the QD versus the applied stresses. Insets: the polar plots of the intensities of the emitted polarized photons from the exciton doublet (blue: the upper level; red: the lower one) of the stressed QD with some specific stresses. In the numerical computation, we consider the $x$-elongated droplet epitaxial GaAs/AlGaAs quantum dot of $\Lambda_x=25.7$nm, $\Lambda_x=17.2$nm, and $H=12$nm. For the model calculation, the length parameters for the spatial extents of the wave function, $l_x=7.9$nm, $l_y=6.7$nm and $l_z=4.3$nm, are taken.}
\label{Fig4}
\end{figure} 

\begin{figure}[t]
\includegraphics[width=15cm]{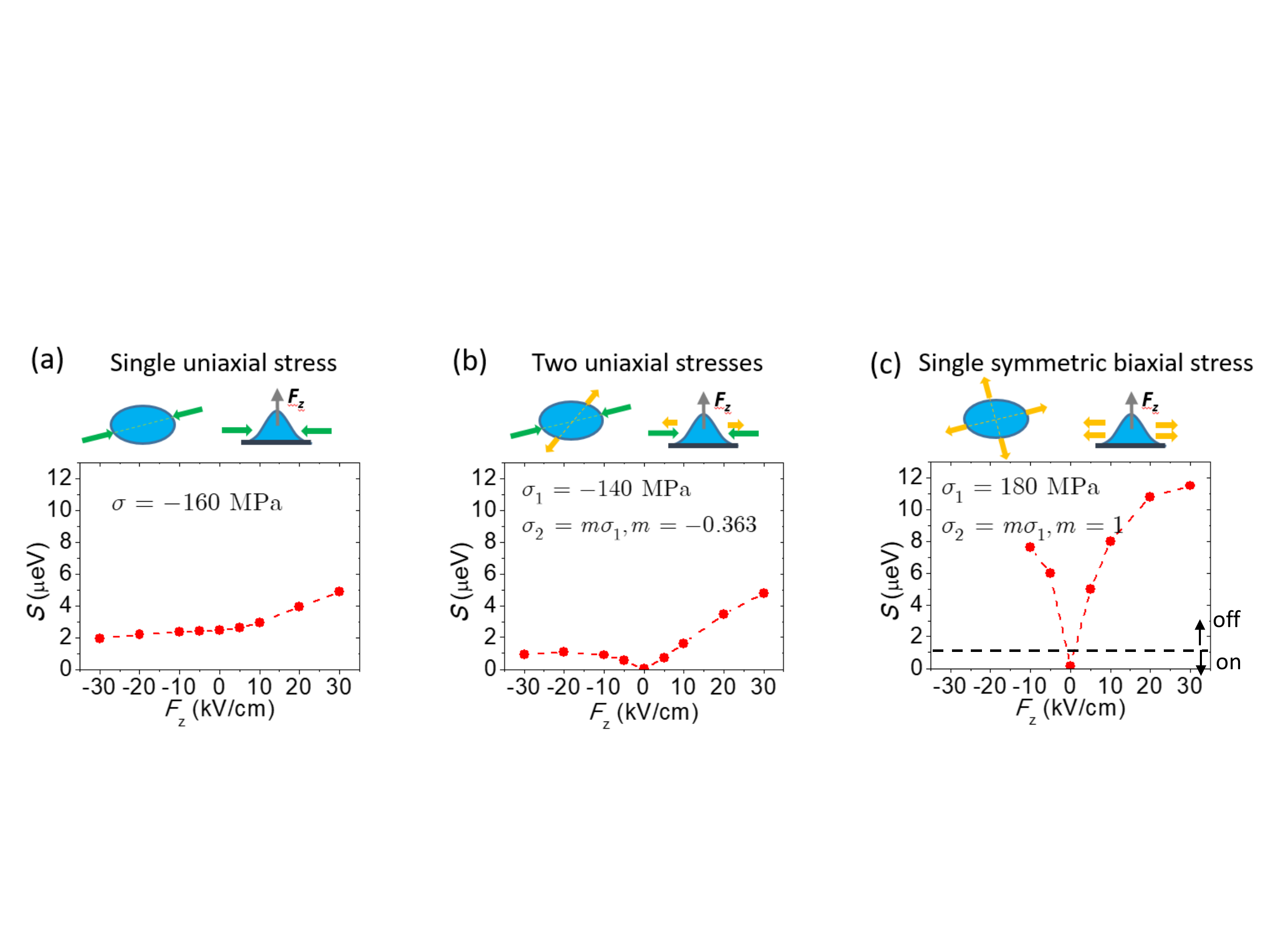}
\caption{
 Calculated excitonic fine structure splittings of the same QD under the stresses considered in Fig.\ref{Fig2} of the main article, and additionally biased by a vertical electrical field, $F_z$. As compared with the cases of single uni-axial stress in (a) and of the dual uni-axial stress in (b), the FSS of the QD controlled by a single symmetric bi-axial stress shown in (c) is most tunable by electric field, and suited to be the "on-demand" electrically triggered EPPE device, which requires the FSS to be electrically switchable below or above the threshold, $S= 1\mu$eV, for the generation of entangled photon pairs. }
\label{Fig5}
\end{figure}

\begin{figure}[t]
\includegraphics[width=18cm]{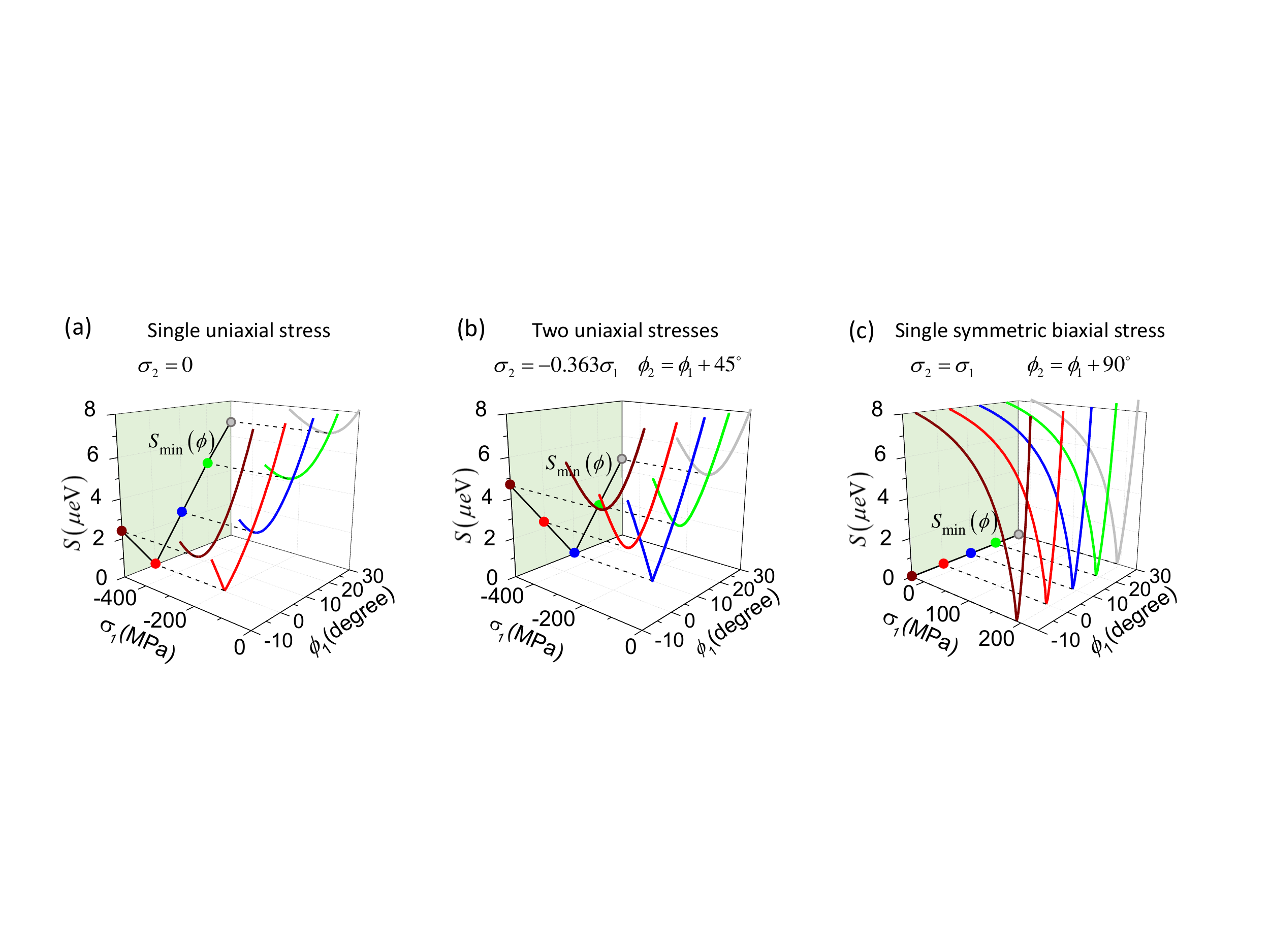}
\caption{
Calculated excitonic fine structure splittings of the same QD considered in Fig.\ref{Fig2} with (a) a single uni-axial stress (b) a set of two uni-axial stresses, and (c) a single symmetric bi-axial stress, versus the varied stress $\sigma_1$ from the actuators arranged in varied orientation, $\phi_1$ (while $m$ and $\Delta\phi_{12}=\phi_{2}-\phi_{1}$ are kept fixed).  The minimal stress-tuned fine structure splittings ($S_{\rm{min}}$) of the stressed QD versus $\phi_1$ are plotted in the x-z plane of the plots.  One sees that, with a single uni-axial stress or the combination of two independently tunable uni-axial stresses, the $S$ can be zero only as $\phi_1$ is at a specific angle. By contrast, with a symmetric bi-axial stress, the $S$ of the stressed QD can be tuned to be stably zero against any variation of $\phi_1$.}
\label{Fig6}
\end{figure}

\section{Theoretical and computational methods}

\subsection{Group theory analysis}
We begin with the group theory analysis for a single exciton in an elongated QD made of zinc-blende
$T_d$ semiconductor. Figure \ref{Fig1} depicts a GaAs QD in the shape of $C_{2v}$ symmetry mounted on a PMN-PT piezoelectricity crystal and stretched by the generated stresses.  Below we summarize the main predicted features of exciton fine structures of the QDs under the stress-control. The technical details of the analysis are given in Sec. S1 of Supplementary Material \cite{supple}

\paragraph*{Bulk} Disregarding the $C_{2v}$ quantum confinement of QD, the states of a spin exciton in a $T_d$ crystal that are created from the direct product of the conduction band and the valence band states are known as $\Gamma_{6c}\times\Gamma_{8v}=\Gamma_{3X}+\Gamma_{4X}+\Gamma_{5X}$,\cite{Singh} which consists of a
doublet in the irreducible representation (irrep.) $\Gamma_{3X}$ and two triplets in $\Gamma_{4X}$ and $\Gamma_{5X}$. The exciton states in the irrep. $\Gamma_{5X}$ ($\Gamma_{3X}$ and $\Gamma_{4X}$) are optically active (inactive) and referred to as the bright exciton (dark exciton) states, according to the Wigner-Eckart theorem. Throughout this work, we shall use the subscript indices $c$, $v$, $X$ and $s$ to indicate conduct band, valence band, exciton state and spin, respectively.

\paragraph*{$C_{2v}$ QDs} With the $C_{2v}$ quantum confinement of QD, the degeneracy of the bright exciton (BX) states in the triplet irrep. $\Gamma_{5X}$ is lifted. 
With the addition of spin-orbit interaction, the conduction band
turns out to be a doublet irrep. $\Gamma_{1c}\times\Gamma_{5s}=\Gamma_{5c}$, and the topmost valence bands $\Gamma_{2v}$ and $\Gamma_{4v}$ are regrouped into the doublet irreps. $\Gamma_{2v}\times\Gamma_{5s} \equiv \Gamma_{5v}^{\left(  2\right)  }$ referred to as the heavy-hole (HH), and
$\Gamma_{4v}\times\Gamma_{5s} \equiv \Gamma_{5v}^{\left(  4\right)  }$ referred to as the light-hole (LH). \cite{Singh}

Since both of the HH and LH bands belong to the same doublet irreps. $\Gamma_{5v}$, it is natural to consider $\Gamma_{5v}^{\prime}=\Gamma_{5v}^{\left(  2\right)}+\tilde{\beta}_{HL}\Gamma_{5v}^{\left(  4\right)  }$ for the valence-band-mixed hole states where the HH- and LH-components are intrinsically mixed. Here, we introduce the complex coefficient $\tilde{\beta}_{HL}\equiv \beta_{HL} e^{-i\phi_{\beta}}$ to parametrize the degree as well as the phase of VBM that are essentially associated with the symmetry of system.
For a $C_{2v}$ QD, $\Gamma_{5v}^{\left(  2\right)}$ and $\Gamma_{5v}^{\left(  4 \right)}$ should keep invariant under the action of the symmetry transformations for $C_{2v}$ ($C_{2z}$, $\sigma_{y}$, $\sigma_{x}$,...) as given in the character table of Fig.~\ref{Fig2}(a), so do $\Gamma_{5v}^{\prime}$. 
Note that, with arbitrary value of $\tilde{\beta}_{HL}$ the mixture of HH and LH, $\Gamma_{5v}^{\prime}=\Gamma_{5v}^{\left(  2\right)
}+\tilde{\beta}_{HL}\Gamma_{5v}^{\left(  4\right)  }$, might not belong to the representation of $C_{2v}$ if the chosen phase angle $\phi_{\beta}$ is improper. In fact, only the phase angles, $\phi_{\beta}=0$ or $\pi$, can match the corresponding symmetry transformations of $C_{2v}$ , and indicates the real value of $\tilde{\beta}_{HL}$. 
A real $\tilde{\beta}_{HL}$ indicates the fixed relative phase between the HH- and LH-components and the fixed orientation of the resulting optical polarizations. Remarkably, the invariance of optical orientation of a single exciton in a QD under the tuning of external fields has been well recognized as a crucial signature to the feasibility of tuning the FSS of the QD down to zero, as will be discussed more later.  \cite{Trotta,Wang,Zhang2015,Plumhof, Chen2016}

\paragraph*{Effects of valence band mixing}

With the VBM nature, the states of the exciton bound by electron-hole Coulomb interactions are created from the direct product of
the $\Gamma_{5c}$ conduction band and the $\Gamma_{5v}^{\prime}$ valence band becomes
$\Gamma_{5c}\times\Gamma_{5v}^{\prime}=\Gamma_{1X}+\Gamma_{2X}+\Gamma
_{3X}+\Gamma_{4X}$, composed of the DX singlet belonging to $\Gamma_{3X}$ and the BX triplet belonging to $\Gamma_{1X}$,
$\Gamma_{2X}$ and $\Gamma_{4X}$, optically polarized along the $z$, $x$ and $y$, respectively. The character table for the spin double group of $C_{2v}$ is presented in Fig.\ref{Fig2}(a). In this work, we are mainly interested in the $x$- and $y$-polarized BX doublet ($\Gamma_{2X}$ and $\Gamma_{4X}$) that can emit light vertically out of the QDs grown on the $(001)$ substrate. 

In principle, the four exciton states ($\Gamma_{1X}, ...,\Gamma_{4X}$) of a $C_{2v}$ QD that belong to the different irreducible representations should own distinctive energies and are subject to level splittings, except that some accidental degeneracy happens. Thus, the only possibility of crossing over the $\Gamma_{2X}$ and $\Gamma_{4X}$ BX levels of a $C_{2v}$ QD is by means of the {\it accidental degeneracy} that might be made by using some external tuning knobs. 
As pointed out previously by Singh and Bester in Ref.\cite{Singh}, the formation of such as accidental degeneracy of the BX doublet of an asymmetric QD could be possible {\it if and only if} the BX states
belong to different irreducible representations. From our analysis, it is shown that, as long as $C_{2v}$ symmetry of QD can be preserved during the FSS-tuning, the BX doublets surely stay in the different irreps. $\Gamma_{2}$ and
$\Gamma_{4}$ (See Fig.\ref{Fig2}(a) and (c) for illustration).

\paragraph*{Effects of external tuning knobs}
Yet, in reality imposing tuning knobs onto a QD to tune the FSS is likely to break the $C_{2v}$ symmetry down to the lower ones, say $C_2$ symmetry.
For a QD in the low symmetry caused by knob-tunings, it turns out that the four exciton states belong to the irrep. $2\Gamma_{1X}+2\Gamma_{2X}$ in the
$C_{2}$ group (See Fig.\ref{Fig2}(b) for the character table) with the two BX singlets optically polarized along the $x$ and $y$ direction belonging to the same irrep. $\Gamma_{2X}$, and it becomes impossible to eliminate the FSS of the QD in any way. Figure~\ref{Fig2}(d) depicts the exciton levels of a QD in the lowest $C_2$ symmetry, which are always anti-crossed and cannot recover the degeneracy by knob-tunings. The same conclusion can be obtained also for the even lower $C_{1}$ symmetry in which only one representation exists. In the situation of the low symmetry, it is thus necessary to use the second
tuning knob to retain the $C_{2v}$ symmetry and the possibility of eliminating the FSS. This accounts for that the successful elimination of the FSSs of QDs usually relies on the use of two tuning knobs, e.g. the combination of two independently controlled stresses or that of a stress and an electric field. \cite{Trotta,Wang,Trotta2016} 

Remarkably, with the recent advances in the micro-machine techniques, the PMN-PT actuators can be fabricated in the multi-legged structures to generate multiple ($N$) stresses (acting as mechanical tuning knobs) allowing for more flexible controls and additional functionalities of devices. Recently, 
the micro-machined piezoelectric actuators in the 3-legged, 4-legged and 6-legged structures have been demonstrated for a full control of the generated in-plane stress tensor in semiconductor nanomembranes.\cite{Trotta2016, Trotta2015, Chen2016, Sanchez} With the multi-legged structure, the total generated stress is composed of multiple stresses each of which can be individually controlled and serves as an independently tuning knob. With the multiple tuning stresses, the EPPE devices that are wavelength-tunable and suited for being quantum repeaters have been successfully fabricated, where two stresses are used for the suppression of the FSS and the others for the tuning of the light wavelength or other functionalities.\cite{Trotta2016, Trotta2015}   

\subsection{Deterministic elimination of exciton FSS by $N$ uni-axial stresses}
The resultant strain in a GaAs QD under $N$ uni-axial stresses with the magnitudes $\{\sigma_i\}$ and the angles with respect to the elongation-axis $\{\phi_i \}$ is derived, through the standard procedures of tensor transformation as detailed in Sec. S2 of Ref.\cite{supple}, as characterized by the non-zero strain tensor elements given by
\begin{eqnarray}
\epsilon_{xx}&=&\dfrac{s_{11}+s_{12}}{2}\cdot (\sum_{i=1}^{N} \sigma_i ) + \dfrac{s_{44}}{4}\cdot (\sum_{i=1}^{N} \sigma_i \cos 2\phi_i ) \notag\\
\epsilon_{yy}&=&\dfrac{s_{11}+s_{12}}{2}\cdot (\sum_{i=1}^{N} \sigma_i ) - \dfrac{s_{44}}{4}\cdot (\sum_{i=1}^{N} \sigma_i \cos 2\phi_i ) \notag\\
\epsilon_{xy}&=&\dfrac{s_{11}-s_{12}}{2} \cdot (\sum_{i=1}^{N} \sigma_i \sin 2\phi_i ) \notag\\
\epsilon_{zz}&=&(s_{11}+s_{12})\cdot (\sum_{i=1}^{N} \sigma_i )
\label{Nstress}
\end{eqnarray}
where the elastic compliance constants are $s_{11}=0.0082$GPa$^{-1}$, $s_{12}=-0.002$GPa$^{-1}$, and $s_{44}=0.0168$GPa$^{-1}$ for GaAs.\cite{Chuang} 

Following the Neumann's principle, the strain tensor given by Eq.(\ref{Nstress}) owns the $C_{2v}$ symmetry if it is invariant under the operation of any $C_{2v}$ symmetry operators. As detailed in Sec. S2.D of Ref.\cite{supple}, one can show that the strain generated by $N$ tuning uni-axial stresses can keep invariant the symmetry of $C_{2v}$ and enables the full FSS-elimination for a $C_{2v}$ QD as long as the following equation for the arrangement of the stresses is fulfilled,
\begin{equation}
\sum_{i=1}^N \sigma_i \sin 2\phi_i=0
\label{S0condit}
\end{equation}
Equation (\ref{S0condit}) can serve as a general guideline for the optimal design of the useful micro-machined actuators that can generate $N$ uni-axial stresses for the deterministic control and elimination of the FSSs of elongated QDs. Without losing the generality, hereafter we shall focus the study on the QDs under single-stress ($N=1$) or dual-stress ($N=2$) controls, which are most feasible to be implemented in experiments.

\subsection{Numerical approaches}

To confirm the prediction of the group theory analysis, we carry out the numerical calculations of the spectral fine structures and optical polarizations of single excitons in GaAs DE-QDs under various single- and dual-stress controls by using the computational methodology employed in Ref.\cite{Taka, Ivchenko,  Kadantsev, Cheng2015}. Considering the GaAs material as a wide band-gapped semiconductor, we neglect the weak coupling between the conduction and valence bands, and compute separately the electronic structures of an electron and a valence hole in a GaAs QD in the single-band theory and the four-band $k\cdot p$ theory, respectively. In the former (latter) theory, the wave function of a conduction electron (a valence hole) in a QD is written as $\psi_{i_e}^e(\vec{r})=g_{i_e}^e(\vec{r}) u_{s_z}^c$ ($\psi_{i_h}^h(\vec{r})=\sum_{j_z=\pm1/2, \pm 3/2 }g_{i_h}^h(\vec{r}) u_{j,j_z}^v$), where $g_i^{e/h}$ are the slowly varying electron/hole envelope functions, $i_{e/h}$ stands for a composite index composed of those of the orbital and spin of an electron/a hole state, $s_z$ is the $z$-component of electron spin, $j_z$ is the $z$-component of the angular momentum $j=3/2$ of valence hole, and $u_{s_z}^{c}$ ($u_{j=3/2,j_z}^{v}$) is the microscopic periodic part of the Bloch function of the conduction (valence) band. Based on the calculated electronic structures of an electron and a valence hole in a GaAs QD, the theory for the electron-hole exchange interaction of an exciton in the QD is established and used to compute the excitonic fine structures.

\subsubsection{Four band $k \cdot p$ model for a valence hole in a stressed QD}

 In the four-band model, the Hamiltonian for a single hole in a stressed QD is formulated as a $4\times4$ matrix composed of the kinetic energy-, strain- and potential parts, $H_{h}=H_k^h+H_{\epsilon}^h+V_{QD}^h$. In the basis of the Bloch functions ordered by $\{u_{j,j_z}\}=\{ |u_{\frac{3}{2},\frac{3}{2}}\rangle, |u_{\frac{3}{2},\frac{1}{2}} \rangle,|u_{\frac{3}{2},-\frac{1}{2}}\rangle,|u_{\frac{3}{2},-\frac{3}{2}}\rangle   \}$, the Hamiltonian is expressed as \cite{Chuang,Liao2012,Cheng2015}
\begin{equation}
 H_h = 
\left(
\begin{array}{cccc}
P+Q & -S & R & 0 \\
-S^+ & P-Q & 0 & R \\
R^+ & 0 & P-Q & S \\
0 & R^+  & S^+ & P+Q
\end{array}
\right)+V_{QD}^hI_{4\times4} \, ,
\label{four}
\end{equation}
where $P=P_k+P_\epsilon$, $Q=Q_k+Q_\epsilon$, $R=R_k+R_\epsilon$, $S=S_k+S_\epsilon$,
$P_k=\frac{\hbar^2\gamma_1}{2m_0}(k_x^2+k_y^2+k_z^2)$, $Q_k=\frac{\hbar^2\gamma_2}{2m_0}(k_x^2+k_y^2-2k_z^2)$,
$R_k=\frac{\sqrt{3}\hbar^2}{2m_0}\left[-\gamma_3(k_x^2-k_y^2)+2i \gamma_2 k_xk_y\right]$,
$S_k=\frac{\sqrt{3}\hbar^2\gamma_3}{2m_0}(k_x-ik_y)k_z$, $P_{\epsilon}=-a_v(\epsilon_{xx}+\epsilon_{yy}+\epsilon_{zz})$, $Q_{\epsilon}=-\frac{b}{2}(\epsilon_{xx}+\epsilon_{yy}-2\epsilon_{zz})$,
$R_{\epsilon}=\frac{d}{2}(\epsilon_{xx}-\epsilon_{yy})-i\sqrt{3}b\epsilon_{xy}$ and $S_{\epsilon}=-\frac{d}{2}(\epsilon_{xz}-i\epsilon_{yz})$, $\vec{k}=(k_x, k_y,k_z) \equiv -i\vec{\bigtriangledown}_{\vec{r}}$ is the wave vector operator,
$\vec{r}=(x,y,z)$ is the coordinate position of carrier, $e>0 (m_0)$ stands for the elementary charge (mass) of free electron, and
$\gamma_1 = 7.1$, $\gamma_2 =2.02$, $\gamma_3 = 2.91$, $a_v=1.16{\rm eV}$, $b =-1.7{\rm eV}$, and $d =-4.55{\rm eV}$ are the Luttinger parameters for GaAs. 
In the numerical studies, according to the observations of atomic force microscope we model the shape of elongated ${\rm GaAs/Al_{0.35}Ga_{0.65}As}$ DE-QDs in terms of the  characteristic function,
\begin{equation}
X(\vec{r})=\left\{\begin{array}{cc}
1, &\quad  0\leq z\leq H{\rm exp}\left(-\frac{x^2}{\Lambda_x^2}-\frac{y^2}{\Lambda_y^2}\right) \\
0, &\quad {\rm elsewhere}
\end{array} \right. \, ,\label{XQD}
\end{equation}
where $H$ is the height of QD  and $\Lambda_{x/y}$ parametrize the lateral characteristic length of QD along the $x/y$ direction.\cite{Keizer, Liao2012}
In this work, we consider asymmetric QDs on $(001)$-substrate and elongated along the crystalline axis of $[1\bar{1}0]$, and specify the growth (elongation) axis as the $z$ ($x$)-axis.
The confining potential of a GaAs/AlGaAs QD for a carrier can be written as $V_{QD}^{\nu}(\vec{r}_{\nu})=V_b^{\nu} \cdot X_{QD}(\vec{r}_{\nu})$, where $\nu=e/h$ denotes electron/hole and the band-offset $V_b^e=300$meV and $V_b^h=200$meV are taken for GaAs/AlGaAs heterostructure.

\subsubsection{Single band model for a conduction electron in a stressed QD}

In the single band model, the Schr\"odinger equations for a single electron in a stressed QD reads
 $H_{e}g_{i_e}^e=E_{i_e}^e g_{i_e}^e$, where 
\begin{equation}
H_e=\frac{\hbar^2 (k_x^2 + k_y^2 + k_z^2)}{2m_e^\ast}+V_{QD}^e(\vec{r}_e)+a_c(\epsilon_{xx}+\epsilon_{yy}+\epsilon_{zz})
\label{singleband}
\end{equation}
is the strain-dependent Hamiltonian for single electron in the single band model, $g_{i_e}^e$ is the envelope wave function of electron
$V_{QD}^e(\vec{r}_e)$ is the position-dependent confining potential for an electron in the dot, $m_e^\ast =0.067m_0$ is the effective mass of electron, $m_0$ is the free electron mass, and  $a_c=-8.013$eV for GaAs.\cite{Chuang} 

In the presence of an electric field, $\vec{F}=(F_x,F_y,_Fz)$, the field-induced Hamiltonian for an electron (a valence hole), $e\vec{F}\cdot \vec{r}_e$ ($-e\vec{F}\cdot \vec{r}_h$), is imposed to Eq.(\ref{singleband}) (Eq.(\ref{four})), where $e \, (>0)$ is the elementary charge of electron. The energy levels and wave functions of a single electron (hole) in a GaAs QD are numerically calculated within the single-band effective mass (four-band $k\cdot p$) theory using the finite-difference method as employed in Ref.\cite{Cheng2015}.

\subsubsection{Computations of the excitonic fine structures of QDs}

Following the methodology of Ref.\cite{Cheng2015}, the Hamiltonian for an interacting exciton in a QD reads 
$ H_X  = \sum_{i_e} E_{i_e}^e c_{i_e}^+c_{i_e}
+ \sum_{i_h} E_{i_h}^h h_{i_h}^+h_{i_h} 
-  \sum_{i_e,j_h,k_h,l_e}V_{i_e,j_h,k_h,l_e}^{eh}c_{i_e}^+h_{j_h}^+h_{k_h}c_{l_e} + \sum_{i_e,j_h,k_h,l_e}V_{i_e,j_h,k_h,l_e}^{eh,xc}c_{i_e}^+h_{j_h}^+h_{k_h}c_{l_e} $, 
where $i_{e}$ ($i_{h}$) represents a composite index composed of the labels of orbital and spin of a single-electron (single-hole) state, $c_{i_e}^+$ and $c_{i_e}$ ($h_{i_h}^+$ and $h_{i_h}$) are the particle creation and annihilation operators, 
\begin{equation}
V_{i_e,j_h,k_h,l_e}^{eh}\equiv \int \int d^3{r_{e}} d^3{r_{h}} \psi_{i_e}^{e\ast}(\vec{r_{1}}) \psi_{j_h}^{h\ast}(\vec{r_{2}}) \frac{e^{2}}{4 \pi \epsilon_0 \epsilon_b |\vec{r}_{12}|}  \psi_{k_h}^{h}(\vec{r_{2}}) \psi_{l_e}^{e}(\vec{r_{1}})
\end{equation}
are the matrix elements of Coulomb interactions causing the electron-hole scatterings, and
\begin{equation}
V_{i_e,j_h,k_h,l_e}^{eh,xc}\equiv \int \int d^3{r_{1}} d^3{r_{2}} \psi_{i_e}^{e\ast}(\vec{r_{2}}) \psi_{j_h}^{h}(\vec{r_{2}}) \frac{e^{2}}{4 \pi \epsilon_0 \epsilon_b |\vec{r}_{12}|}  \psi_{k_h}^{h \ast}(\vec{r_{1}}) \psi_{l_e}^{e}(\vec{r_{1}}) 
\end{equation}
are those of {\it e-h exchange} interactions, $\vec{r}_i$ denotes the coordinate position of particle, $\vec{r}_{12}\equiv \vec{r}_1- \vec{r}_2$, $\epsilon_0$ is vacuum permittivity, $\epsilon_b$ is the dielectric constant of QD material ($\epsilon_b =12.9$ for GaAs), $E_{i_e}^e$ and $E_{i_h}^h$ ($\psi_{i_e}^{e}$  and $\psi_{i_h}^{h}$) are the eigen energies (wave functions) of a single electron and single hole in the QD, respectively.

Since our interest is in the fine structures of the lowest exciton states, we take into account only the relevant lowest single-electron and single-hole orbitals and, for the brevity of notation, label them only with the spin indices, i.e. $|\psi^e_{i_e=\uparrow_e/\downarrow_e} \rangle \equiv |\uparrow_e/\downarrow_e \rangle $, ($|\psi^h_{i_h=\Uparrow_h'/\Downarrow_h'} \rangle \equiv |\Uparrow_h'/\Downarrow_h' \rangle$), where $\uparrow_e/\downarrow_e$ denotes the up/down electron spin and $\Uparrow_h'/\Downarrow_h'$ indicates the up/down pseudo-spin of a HH-like hole state. 
In the reduced basis of the direct products of the single-electron and -hole states, $|\uparrow_e\rangle |\Downarrow_h'\rangle$ and $|\downarrow_e \rangle |\Uparrow_h'\rangle$, being the two lowest BX configurations, the Hamiltonian for an valence-band-mixed bright exciton (BX) in a QD is written as a $2\times 2$ matrix,  

\begin{eqnarray}\label{hamilX}
H_X= 
\left(
       \begin{array}{cc}
E_X^{(0)} & \tilde{\Delta}_{eff}^{xc} \\
\tilde{\Delta}_{eff}^{xc \, \ast} & E_X^{(0)} \\
       \end{array}
     \right) \, ,
     \label{Hx}
\end{eqnarray}
where $E_X^{(0)}=E_{\uparrow_e}^e+E_{\Downarrow_h'}^h -V_{\uparrow_e\Downarrow_h'\Downarrow_h'\uparrow_e}^{eh} = E_{\downarrow_e}^e+E_{\Uparrow_h'}^h -V_{\downarrow_e\Uparrow_h'\Uparrow_h'\downarrow_e}^{eh}$ denotes the energy of exciton regardless of the $\it e-h$ exchange interactions, and $\tilde{\Delta}_{eff}^{xc}\equiv V_{\uparrow \Downarrow' \Uparrow' \downarrow }^{ehxc}$ is the off-diagonal matrix element of the $\it e-h$ exchange interaction that couples the two VBM bright exciton configurations of opposite angular momenta and results in the FSS of the exciton doublet, $|S|=2|\tilde{\Delta}_{eff}^{xc}|$.

In the numerical calculation, the matrix elements of {\it e-h} exchange interactions are divided by the short-ranged and long-ranged parts according to the averaged Wigner-Seitz radius, and computed separately.\cite{Cheng2015, Kadantsev} The former is treated in the dipole-dipole interaction approximation and numerically integrated using trapezoidal rules and graphics processing unit (GPU) parallel computing technique for numerical acceleration.  The latter is considered for the matrix elements involving the exciton basis of same angular momenta and evaluated using the formalism of Eq.(2.17) in Ref.\cite{Taka}, in terms of the energy splitting between the bright- (BX) and dark-exciton (DX) states of a QD, $E_X^S = \Delta_{eh,bulk}^{xc} \times [ \pi (a_B^\ast)^3 \int d^3r |g_{s_z=\pm\frac{1}{2}}^e|^2 |g_{j_z=\mp\frac{3}{2}}^h|^2 ]$, which is extrapolated, in terms of the effective Bohr radius of exciton $a_B^\ast$, from the BX-DX splitting $\Delta_{eh,bulk}^{xc}=20\mu$eV of a pure HH-exciton in the GaAs bulk. 

From the solved eigen states, $|\Psi_{\pm}^X \rangle$, and the corresponding eigen energies, $E_{\pm}^X= E_X^{(0)} \pm |\tilde{\Delta}_{eff}^{xc}|$, for Eq.(\ref{Hx}), one can calculate the intensities $I_{\pm}(\hat{e}, \omega)$ of the $\hat{e}$-polarized photo-luminescences (PLs) from the exciton states using the formalism of the Fermi's golden rule, as employed in Ref.\cite{Cheng2015}.  
The PL intensity as a function of the polarization $\hat{e}$ reaches the maximum, $I_{\pm, max}=I_{\pm}(\hat{e}=\hat{e}_{\pm}, \omega= E_{\pm}^X/\hbar) $, as the polarization is along the optical axis of the exciton state $|\Psi_{\pm}^X \rangle$, specified by the unit vector along the axis, $\hat{e}_{\pm}=(\cos \phi_{\pm}^{\rm{opt}}, \sin \phi_{\pm}^{\rm{opt}}, 0)$. Note that the both exciton basis for Eq.(\ref{Hx}) are circularly polarized. The {\it e-h} exchange interaction $\tilde{\Delta}_{eff}^{xc}$ leading to the off-diagonal matrix element of Eq.(\ref{Hx}) mixes the both circularly polarized exciton basis and the resulting eigen states of exciton usually turn out to be linear polarized. As dicussed thoroughly in Ref. \cite{Cheng2015}, the exciton eigen states of an $x$-elongated QD are polarized in the direction parallel or perpendicular to the elongation of the QD  (along the $x$- or $y$-axes) as the off-diagonal matrix elements are real. On the other hand, the misaligned polarization of a exciton eigen state from the elongation axis results from the off-diagonal matrix elements that are complex and can be characterized by a non-zero phase angle, which is related to the phase angle $\phi_{\beta}$ introduced previously for the HH-LH coupling of an exciton in the group theory analysis and is an indication of the lowered symmetry. As previously discussed in the group theory analysis, the lowering of the symmetry of QD makes it no longer possible to tune the FSS of an elongated QD down to zero. Thus, the feasibility of using external tuning knobs to eliminate the FSS of an elongated QD can be observed from the orientation, i.e. $\phi_{\pm}^{\rm{opt}}$, of the optical polarization of the exciton states (see if it is aligned to or misaligned from the elongation axis). More discussion on the issue for the specific examples of QDs will be given in the next section.   

\section{Results and discussion}

\subsection{Useful stress actuators predicted by the group theory}

\subsubsection{Single uni-axial stress}
As a known example, using a single uni-axial stress can fully eliminate the FSS of an elongated QD only if the stress- and elongation-axes are {\it exactly} aligned ($\phi_1=0^\circ$).\cite{Singh} Substituting $\sigma_2=0$ into Eq.(\ref{S0condit}), we obtain $\sigma_1 \sin 2\phi_1 =0$, indicating $\phi_1=0$ for $\sigma_1 \neq 0$, i.e. the perfect alignment of the stress $\sigma_1$ onto the elongation axis of QD. 
By contrast, with a misaligned single uni-axial stress of $\phi_1\neq 0$, Eq.(\ref{S0condit}) is no longer fulfilled unless $\sigma_2\neq 0$, indicating the need of the second tuning knob for retaining the $C_{2v}$ symmetry of QD and the possibility of fully eliminating the FSS of the dot. 

\subsubsection{Two uni-axial stresses}
For the cases of two uni-axial stresses ($\sigma_1\neq 0,\sigma_2\neq 0$), let us define the stress ratio by $m \equiv \frac{\sigma_2}{\sigma_1}$ and rewrite Eq. (\ref{S0condit}) as  $m (\phi_1,\phi_2) = -\frac{\sin 2\phi_1}{\sin 2\phi_2}$ for further analysis. In Fig. \ref{Fig3}, we plot the contour curves of $m$ as a function of $\phi_1$ and $\phi_2$. 
By tracing the $m$-contours in Fig.\ref{Fig3} where Eq.(\ref{S0condit}) is surely fulfilled, we are able to predict various useful dual-stress actuators that can promisingly generate the strain remaining in the $C_{2v}$ symmetry and enable the FSS-elimination. Accordingly, one can determine the strength ratio $m$ of a dual-stress arranged in specific axes with fixed $(\phi_1,\phi_2)$ for the FSS-elimination. In turn, for a stress-actuator that can generate a pair of stresses with a fixed strength ratio $m$, Fig. \ref{Fig3} guides us to find the optimal arrangement of $\phi_1$ and $\phi_2$ for the purpose of FSS-elimination.

For instance, a single uni-axial stress perfectly aligned to the elongation axis is represented by the vertical line for $m=0$ and depicted in the inset (a) in Fig.\ref{Fig3}, which is useful to eliminate the FSS of an $x$-elongated QD as discussed previously. 
The horizontal contour line at $\phi_2=90^{\circ}$ labelled by $m \rightarrow \infty$ in Fig.\ref{Fig3} indicates a single uni-axial stress ($\sigma_1=0,\sigma_2\neq 0$) that is perpendicular to the elongation axis, as depicted by inset (d). 
In another case, the black circle at $(\phi_1, \phi_2)=(0^{\circ}, 90^{\circ})$  that connects all $m$ contour curves in Fig.\ref{Fig3} represent a generic orthogonal bi-axial stress with the freely tuned $\sigma_1$ and $\sigma_2$ as depicted in inset(e). A feasible example of inset(e) is the asymmetric bi-axial stress produced from the $(100)$ facet of PMN-PT crystal, which has been successfully employed to tune and suppress the FSSs of elongated InGaAs SK-QDs, yet, with the strict requirement for the precise alignment of the stress- and elongation axes.\cite{Zhang2015, Chen2016}
  
If the $\sigma_1$-axis is misaligned from the elongation one ($\phi_1 \neq 0$), by tracing the vertical dashed line of $\phi_1$ and examining the $m$-values of the crossed contours by the vertical line, one can find that all of the stress-ratios  required for FSS-elimination are non-zero, i.e. $ m =\sigma_2/\sigma_1\neq 0$. This indicates that a second tuning stress ($\sigma_2 \neq 0$) is necessary if the uni-axial stress axes cannot be aligned to the elongation one. As a specific example, for a set of two mis-aligned uni-axial stresses with $\phi_1= 10^\circ$ and $\phi_2=55^\circ$, the stress-ratio $m=-0.363$ is predicted to suppress the FSS of a QD with the stress (See the inset (b) of Fig.~\ref{Fig3}). The numerical confirmation for those predictions is presented in the next section. 

\subsubsection{Deterministic FSS-elimination with a single stress: beyond the two-tuning-knob scheme}
Beyond the use of two tuning knobs, a single knob tuning for making $S=0$ is possible if some underlying relationship between the tuning stresses exists and can be utilized to reduce the number of independent variables of Eq.(\ref{S0condit}). Among the predicted useful dual-stress actuators, we find that a symmetric bi-axial stress ($m=1$ and $\phi_2=\phi_1 + 90^\circ$) can act as a single mechanical tuning knob, represented by the red straight line of $m=1$ and the inset (f) in Fig.\ref{Fig3}. Besides the equality of $\sigma_1$ and $\sigma_2$, the straightness of the $m=1$ contour indicates the fixed angle between the $\sigma_1$- and $\sigma_2$-axes, $\Delta \phi_{21}=\phi_2-\phi_1=90^\circ$, which can be naturally kept by the cubic nature of the crystal structure of PMN-PT piezoelectricity crystal.  
Such a symmetric bi-axial stress can be generated naturally from the $(001)$ facet of PMN-PT crystal under a {\it single} tuning electrical bias and remain invariant in the symmetry no matter how orientated the PMN-PT crystal is.

\subsection{Numerical results}

\subsubsection{Single uni-axial stress}
Figure~\ref{Fig4}(a) shows the numerically calculated FSS ($S$) between the lowest bright exciton states of the $x$-elongated GaAs DE-QD of $H=12$nm, $\Lambda_x=26$nm and $\Lambda_y=17$nm under a single tuning uni-axial stress along the direction with the angle $\phi_1=10^{\circ}$ with respect to the $x$-axis. As we expected, the application of the mis-aligned uni-axial stress leads to the reduction of the symmetry of the QD down to $C_{1}$ and cannot fully eliminate the FSS.\cite{Seidl,Singh,Zhang2015}  To retain the $C_{2v}$ symmetry of QD, one can introduce and use a second tuning knob.
 
\subsubsection{Two uni-axial stresses}
Figure~\ref{Fig4}(b) shows the calculated FSS of the same stressed QD with, additionally, a second uniaxial stress set in the fixed direction with $\phi_2=55^{\circ}$. The FSS of the QD tuned by the two mechanical knobs is shown fully eliminated with $\sigma_1=-140$MPa and $\sigma_2=51$MPa, whose ratio $m=\frac{\sigma_2}{\sigma_1}=-0.363$ is exactly as predicted by Eq.(\ref{S0condit}). Intuitively, the necessity of using two tuning knobs to make FSS-elimination is widely understood from the observed correlation between the FSS and the optical anisotropy featured by the degree as well as orientation of polarization,\cite{Trotta,Wang,Zhang2015,Plumhof, Chen2016} as evidenced here by the comparison between Fig.~\ref{Fig4}(a)-(c) and Fig.~\ref{Fig4}(d)-(f).  Figure~\ref{Fig4}(d)-(f) present the calculated angles, $\phi_{+}^{\rm{opt}}$, of the optical axes for the upper exciton level of the QD under the different types of stresses, corresponding to Fig.~\ref{Fig4}(a)-(c), respectively.  Note that, once upon the FSS of a QD can be tuned to be zero (See Fig.~\ref{Fig4} (b),(c),(e) and (f)), the orientation of optical polarization remains unchanged against the stress-tuning (except that $S=0$ happens). 
The angle-invariance of the optical polarization implies the preservation of the $C_{2v}$ symmetry of QD. 
From the above observations, the use of two tuning knobs is effective to tune the FSS and simultaneously keep the orientation of polarization invariant. \cite{Trotta}
 
\subsubsection{Deterministic FSS-elimination with a single stress}
Figure~\ref{Fig4}(c) shows the numerically calculated results for the $x$-elongated QD under a {\it single} symmetric bi-axial stress with the misaligned axes from the $x$-and $y$-axes by $\phi_1=\phi_2 - 90^\circ=10^{\circ}$. As predicted by previous analysis, the numerical simulation confirms that the excitonic FSS of the $C_{2v}$ QD indeed can be eliminated fully by the {\it single} tuning bi-axial stress of $\sigma = 177$MPa, in spite of the misalignment of the stress and elongation axes. This result examples that, beyond common intuitive understanding, the use of two tuning knobs is a sufficient but not always a necessary condition for eliminating the FSS of a QD. The full elimination of the FSS of the DE-QD with the single mechanical tuning knob is achieved by taking the advantage of the compatibility of crystal symmetry between the QD- and piezoelectricity actuator materials, which can always ensure the $C_{2v}$ preservation and allow for some accidental degeneracy happening in the BX doublet.  
 The use of a {\it symmetric} bi-axial stress for tuning the FSSs of self-assembled QDs has been previously explored but was not found so advantageous in the FSS-elimination for the studied InGaAs SK-QDs. \cite{Wang}   The usefulness of symmetric bi-axial stress is limited for elongated InGaAs SK-QDs since there exist intrinsic strains in the SK-QDs, which are themselves {\it asymmetric} and spoil the symmetry of the applied bi-axial stresses. \cite{Wang}

\subsubsection{Electrical tunabilities of stressed QDs }

The high tunability for the FSSs of QDs is a crucial functional feature for realizing the "on-demand" QD-based entangled photon pair emitters that requires the efficient switch-on ($S<1\mu$eV) and -off ($S \gg 1\mu$eV) of the devices by electrically gating for the integrated application with micro-electronics. Figure~\ref{Fig5} presents the numerically calculated excitonic fine structure splittings of the stressed QD as considered in Fig.\ref{Fig4}(a)-(c) and additionally applied by a vertical electrical field, $\vec{F}=(0,0,F_z)$. As the FSS of the QD is tuned to be nearly vanishing by an appropriate stress, an external electric field is used here to re-open the splitting to switch-off the device. In turn, the device can be switched on by turning off the applied electric field. The FSS of the QD under the three types of stress-control is shown all electrically tunable, but to different extents. Among them, only the FSS of the QD imposed by a single symmetric bi-axial stress (Fig.\ref{Fig5}(c)) is so well tunable by the external electric field that the FSS can be varied over a practically useful wide range of energy. In Fig.\ref{Fig4}(c), the FSS of the stressed QD is shown quickly changed from $S=0$ to over $5\mu$eV by applying the small electric field $F_z\sim 5$kV/cm onto the QD. This is attributed to the non-linear nature of the bi-axial term (that will be discussed more by the model analysis later) in the VBM of an exciton confined in the stressed QD, which can make more impact on the VBM-relevant FSS of the QD as the wave function extents are varied only slightly by an external electric field. 

\section{Effective exciton model}

Following the methodology in Ref.~\cite{Cheng2015}, we proceed to establish a simplified generic exciton model for stress-controlled QDs that is in consistency with the previous analysis and numerical results, and allows for more physical analysis.
By treating the HH-LH coupling terms in the four-band theory as perturbation, one can derive an effective exciton Hamiltonian in the compact form of $2\times 2$ matrix, explicitly in terms of the QD parameters and applied stresses, as presented below.

 In the lowest-order approximation, the lowest spin-up (spin-down) HH-like hole state, $\vert \Uparrow_h'\rangle$ ($\vert \Downarrow_h'\rangle $), of a stressed QD can be written as $\vert \Uparrow_h'\rangle  \approx \vert \Uparrow_h\rangle - \tilde{\beta}_{HL}\vert \downarrow_h \rangle$ ($\vert \Downarrow_h'\rangle  \approx \vert \Downarrow_h\rangle - {\tilde{\beta}_{HL}}^\ast\vert \uparrow_h \rangle$), composed of the dominant pure HH component, $\vert \Uparrow_h\rangle$ ($\vert \Downarrow_h\rangle$) mixed by the secondary LH one, $\vert \uparrow_h\rangle$ ($\vert \downarrow_h \rangle$), via the complex coefficient, $\tilde{\beta}_{HL} \equiv \beta_{HL} e^{ -i\phi_{\beta}}$ that reflects the degree of VBM. \cite{Leger}
In the parabolic model, the envelop wave function of the HH component is modelled by $\langle \vec{r} \vert \Uparrow_h'\rangle \equiv \psi_{\Uparrow_h'}(\vec{r}) \approx \phi_{0}(\vec{r}) u_{j_z=3/2}(\vec{r}) -\beta_{HL}^\ast \phi_{0}(\vec{r}) u_{j_z=-1/2}(\vec{r})$, where $\phi_{0}(\vec{r})=\sqrt{\frac{1}{\pi^{3/2}l_x l_y l_z}}\exp\left\{-\frac{1}{2}\left[\left(\frac{x}{l_x}\right)^2
+\left(\frac{y}{l_y}\right)^2+\left(\frac{z}{l_z}\right)^2\right]\right\}$ is the wave function of the lowest Fock-Darwin state in the parabolic model and $l_{\alpha=x,y,z}$ is the characteristic length of the wave function extent along the $\alpha$-direction.
In the basis of the VBM-exciton configurations, $\frac{1}{\sqrt{2}}(|\downarrow_e \Uparrow_h'\rangle \pm |\uparrow_e \Downarrow_h'\rangle$, one can derive the effective exciton Hamiltonian for a stressed QD as
\begin{eqnarray}
H_X'& =& 
\left(
\begin{array}{cc}
E_{X}^{(0)}+\Re[\tilde{\Delta}_{eff}] &\Im[\tilde{\Delta}_{eff}]   \\
-\Im[\tilde{\Delta}_{eff}] &  E_{X}^{(0)}-\Re[\tilde{\Delta}_{eff}]
\end{array}
\right) \, .
\label{twobytwo}
\end{eqnarray}
where $E_{X}^{(0)}$ denotes the average energy of the spin-split exciton levels, and $\tilde{\Delta}_{eff}$ is the complex matrix element of the {\it e-h} exchange interaction between the two VBM-exciton configurations, $|\downarrow_e \Uparrow_h'\rangle$ and $|\uparrow_e \Downarrow_h'\rangle$, which, following Ref.\cite{Cheng2015}, can be formulated as 
\begin{eqnarray}
\widetilde{\Delta}_{eff} =-\Delta_1+ \dfrac{2}{\sqrt{3}}E^S_X\cdot\widetilde{\beta}_{HL} \, 
\label{Deltaeff}
\end{eqnarray}
where $-\Delta_1$ is the matrix element of the attractive long-ranged {\it e-h} exchange interaction between the pure-HH exciton, $|\downarrow_e \Uparrow_h\rangle$ and $|\uparrow_e \Downarrow_h\rangle$,
the second term on the RHS originates from the short-ranged {\it e-h exchange} interaction associated with the VBM, $\widetilde{\beta}_{HL}$, and $E^S_X$ is the exchange splitting between the BX and DX states of the QD that, as an empirical parmeter, can be extrapolated from the measured BX-DX splitting of GaAs bulk, $E_S^{X,bulk}=20\mu$eV for GaAs.\cite{Cheng2015} 
As a result, the fine structure splitting of exciton is given by
\begin{eqnarray}
S&=&2\sqrt{(\Im[\tilde{\Delta}_{eff}])^2+(\Re[\tilde{\Delta}_{eff}])^2} 
\label{Splitting}
\end{eqnarray}

From the solved eigen states of Eq.\ref{twobytwo}, the polarized PL can be calculated using the formalisms based on the Feri's golden rules as presented in Ref.\cite{Cheng2015}. 
For a QD with two uni-axial stresses, the VBM coefficient is derived as 
\begin{equation}
\widetilde{\beta}_{HL}=\dfrac{\rho_{HL}^0+\lambda(\sigma_{1}e^{2i\phi_1}+\sigma_{2}e^{2i\phi_2})}{\Delta_{HL}^0+\mu \sigma_b }\, ,
\label{betaHL}
\end{equation}
where $\sigma_b\equiv \sigma_1 +\sigma_2$ is the bi-axial stress,  $\rho_{HL}^0$ ($\Delta_{HL}^0$) is the matrix element of HH-LH coupling (energy difference between the pure HH and LH levels) in the absence of strain,  and the constants, $\lambda$ and $\mu$, are associated with the deformation parameters of QD.  
Equation (\ref{Splitting}) shows that $S=0$ requires that the both of real and imaginary parts of the exchange interaction vanish, i.e. $\Im[\tilde{\Delta}_{eff}]=0$ and $\Re[\tilde{\Delta}_{eff}]=0$. From Eqs. (\ref{Deltaeff}) and (\ref{betaHL}), the former condition,  $\Im[\tilde{\Delta}_{eff}]=0 \propto \Im[ \tilde{\beta}_{HL}] =0$ leads to that $\tilde{\beta}_{HL}$ is real and $\sigma_1\sin2\phi_1 +\sigma_2\sin2\phi_2 =0$, same as Eq.(\ref{S0condit}) derived by the group theory.

\subsection{Advantageous effects of a symmetric bi-axial stress: the model analysis}
With the proper consideration of the strain-dependent VBM on the base of the four-band theory, in Eq.(\ref{betaHL}) the VBM (parametrized by $\widetilde{\beta}_{HL}$) in a QD-confined exciton is shown tunable by uni-axial stresses or a bi-axial one, so are the FSS ($S$) according to Eqs.(\ref{Deltaeff}) and (\ref{Splitting}). In Eq.(\ref{betaHL}), one sees that the terms of uni-axial stress appear in the numerator and are associated with the orientations of the stress axes ($\phi_1$ and $\phi_2$). In contrast, that of biaxial stress lying in the denominator is shown irrelevant to the orientation angles of the applied uni-axial stresses. This implies that tuning the FSS of a QD with a single bi-axial stress could be free from uncertainty in the poor-controlled alignment of the stress and elongation axes, while the orientation alignment of uni-axial stress and elongation axes is critical for the FSS-tuning.

For the comparison with the numerical results in Fig.\ref{Fig4}, the FSSs and the optical polarizations of the stressed QD are calculated by using the exciton model with the length parameters of the wave function, $l_x=7.9$nm, $l_y=6.7$nm and $l_z=4.3$nm, as presented by the black solid lines in Fig.\ref{Fig4}, showing the consistence with the numerical results.   
Our derived exciton model is consistent with and even beyond the previous one in Ref.\cite{Gong} that was developed for their studies of strained InGaAs/AlGaAs SK-QDs. In the latter model, the matrix elements of the exciton Hamiltonian were assumed to be linearized with respect to the applied stress and the bi-axial term in the denominator of Eq.(\ref{betaHL}) was overlooked. The assumption of linear stress-dependence is acceptable for the studied InGaAs/AlGaAs SK-QDs where the intrinsic bi-axial strain from the lattice mismatch between InGaAs and AlGaAs is significant so that $\Delta_{HL}^0 \gg \mu \sigma_b$ and the effect of external bi-axial stress, $\sigma_b$, is thus negligible. For un-strained GaAs DE-QDs under our study, $\Delta_{HL}^0 $ is small because of the absence of intrinsic bi-axial strain and the application of an external bi-axial stress makes more impaction onto the electronic and excitonic structures. Expanding Eq.(\ref{betaHL}) in terms of stress, an external bi-axial stress lying in the denominator yields considerable high-order stress terms, making the FSSs of DE-QDs sensitive to and well tunable by external stresses as shown in Fig.~\ref{Fig4}(c) . For the same sake, the non-linear nature of the bi-axial term in the denominator of Eq.(\ref{betaHL}) for the VBM of exciton leads to the high electrical tunability of the FSS of the QD with a bi-axial stress, as presented in Fig.~\ref{Fig5}.
Figure \ref{Fig4}(c) reveals the pronounced effect of bi-axial stress on the high tunability (maximally $\sim 600\mu \rm{eV/GPa}$) for the FSS of the QD, which might be related to the stress-enhanced super-coupling in the valence band mixing (VBM) of exciton as recently reported by Ref.~\cite{Luo}.

Figure \ref{Fig6} shows the calculated FSS, $S$, of the stressed QD versus the stresses generated from the micro-machined PMN-PT actuators of Fig.~\ref{Fig4} but arranged in various orientations.  The minimum stress-tuned FSS's, $S_{\rm{min}}$, of the stressed QD versus the varied orientations of the stress actuators are projected onto the $x-z$ plane of Fig.~\ref{Fig6}.    
Figure \ref{Fig6} (a) and (b) show that it is very critical to set the orientations of a single uni-axial stress or a set of two asymmetric uni-axial ones so as to fully eliminate the FSS of the stressed QD. With the application of the single uni-axial stress (the set of two uni-axial stresses), $S$ can be fully vanishing only as the orientation of the stress of $\sigma_1$ is set exactly to be $\phi_1=0^\circ$ ($\phi_1=10^\circ$). By contrast, the $S_{\rm{min}}$ of a QD can be eliminated always by a tuning symmetric bi-axial stress, disregarding any orientation variation of the bi-axial stress actuator, as shown in Fig.~\ref{Fig6}(c).  It is the phase-irrelevance of the bi-axial stress term in the denominator of Eq.(\ref{betaHL}) that makes the robustness of the FSS-tuning against the variation of actuator orientation.

\section{Conclusion}

In summary, we present a theoretical and computational investigation of the excitonic FSSs of GaAs/AlGaAs DE-QDs mechanically tuned by the stress actuators of micro-machined PMN-PT crystals. 
From group theory analysis confirmed by fully numerical simulation, we reveal the general principle for the optimal arrangement of two uni-axial stresses whose application onto elongated an elongated GaAs DE-QD can certainly eliminate the FSS of an exciton therein. Moreover, as a main finding of this work, we point out that the use of two tuning knobs for a certain elimination of the FSSs of QDs is a sufficient but not always a necessary condition as commonly believed.  
 The feasibility of using only single knob for making QD-based EPPE devices is significant to simplify the process of device fabrication and allows for developing versatile photonic devices crucial in integrated photonic systems. As a feasible example, a {\it single} symmetric bi-axial stress naturally generated from the $(001)$ PMN-PT actuator can be a single tuning knob for eliminating the FSSs of DE-QDs, whose feasibility is achieved by taking the advantage of the compatibility of the crystal structure symmetries of the QD- and piezoelectricity materials.  Beyond the existing exciton models for stress-tuned QDs, our derived effective exciton model on the base of multi-band theory well captures the non-linearity nature of bi-axial stress terms in the exciton Hamiltonian and enables us to understand more the usefulness of symmetric bi-axial stress, including the robustness of deterministic FSS-elimination against the poor-controlled orientations of stress actuators and the high electrical and mechanical FSS-tunability crucial for realizing "on-demand" electrically triggered EPPEs. 
 
\section{Acknowledgements}
The study is supported by the Ministry of Science and Technology, Taiwan, under contracts, MOST-106-2112-M-009-015-MY3, MOST-106-2221-E-009-113-MY3, and MOST-107-2633-E-009-003, and by National Center for High-performance Computing (NCHC), Taiwan.

\newpage
\clearpage
\bibliography{Biaxial_FSS_QD_refs}

\end{document}